\documentstyle[twoside]{article}

\catcode`\@=11
\long\def\@makefntext#1{
\protect\noindent \hbox to 3.2pt {\hskip-.9pt  
$^{{\eightrm\@thefnmark}}$\hfil}#1\hfill}		

\def\thefootnote{\fnsymbol{footnote}}
\def\@makefnmark{\hbox to 0pt{$^{\@thefnmark}$\hss}}	
	
\def\ps@myheadings{\let\@mkboth\@gobbletwo
\def\@oddhead{\hbox{}
\rightmark\hfil\eightrm\thepage}   
\def\@oddfoot{}\def\@evenhead{\eightrm\thepage\hfil
\leftmark\hbox{}}\def\@evenfoot{}
\def\sectionmark##1{}\def\subsectionmark##1{}}

\oddsidemargin=\evensidemargin
\renewcommand{\thefootnote}{\fnsymbol{footnote}}

\newcounter{sectionc}\newcounter{subsectionc}\newcounter{subsubsectionc}
\renewcommand{\section}[1] {\vspace{12pt}\addtocounter{sectionc}{1} 
\setcounter{subsectionc}{0}\setcounter{subsubsectionc}{0}\noindent 
	{\tenbf\thesectionc. #1}\par\vspace{5pt}}
\renewcommand{\subsection}[1] {\vspace{12pt}\addtocounter{subsectionc}{1} 
	\setcounter{subsubsectionc}{0}\noindent 
	{\bf\thesectionc.\thesubsectionc. {\kern1pt \bfit #1}}\par\vspace{5pt}}
\renewcommand{\subsubsection}[1] {\vspace{12pt}\addtocounter{subsubsectionc}{1}
	\noindent{\tenrm\thesectionc.\thesubsectionc.\thesubsubsectionc.
	{\kern1pt \tenit #1}}\par\vspace{5pt}}

\newcounter{appendixc}
\newcounter{subappendixc}[appendixc]
\newcounter{subsubappendixc}[subappendixc]
\renewcommand{\thesubappendixc}{\Alph{appendixc}.\arabic{subappendixc}}
\renewcommand{\thesubsubappendixc}
	{\Alph{appendixc}.\arabic{subappendixc}.\arabic{subsubappendixc}}

\renewcommand{\appendix}[1] {\vspace{12pt}
        \refstepcounter{appendixc}
        \setcounter{figure}{0}
        \setcounter{table}{0}
        \setcounter{lemma}{0}
        \setcounter{theorem}{0}
        \setcounter{corollary}{0}
        \setcounter{definition}{0}
        \setcounter{equation}{0}
        \renewcommand{\thefigure}{\Alph{appendixc}.\arabic{figure}}
        \renewcommand{\thetable}{\Alph{appendixc}.\arabic{table}}
        \renewcommand{\theappendixc}{\Alph{appendixc}}
        \renewcommand{\thelemma}{\Alph{appendixc}.\arabic{lemma}}
        \renewcommand{\thetheorem}{\Alph{appendixc}.\arabic{theorem}}
        \renewcommand{\thedefinition}{\Alph{appendixc}.\arabic{definition}}
        \renewcommand{\thecorollary}{\Alph{appendixc}.\arabic{corollary}}
        \renewcommand{\theequation}{\Alph{appendixc}.\arabic{equation}}
        \noindent{\tenbf Appendix \theappendixc #1}\par\vspace{5pt}}
\newcommand{\subappendix}[1] {\vspace{12pt}
        \refstepcounter{subappendixc}
        \noindent{\bf Appendix \thesubappendixc. {\kern1pt \bfit #1}}
	\par\vspace{5pt}}
\newcommand{\subsubappendix}[1] {\vspace{12pt}
        \refstepcounter{subsubappendixc}
        \noindent{\rm Appendix \thesubsubappendixc. {\kern1pt \tenit #1}}
	\par\vspace{5pt}}

\newcommand{\textlineskip}{\baselineskip=13pt}
\newcommand{\smalllineskip}{\baselineskip=10pt}

\def\eightcirc{
\begin{picture}(0,0)
\put(4.4,1.8){\circle{6.5}}
\end{picture}}
\def\eightcopyright{\eightcirc\kern2.7pt\hbox{\eightrm c}} 

\newcommand{\copyrightheading}[1]
	{\vspace*{-2.5cm}\smalllineskip{\flushleft
	{\footnotesize International Journal of Modern Physics A, #1}\\
	{\footnotesize $\eightcopyright$\, World Scientific Publishing
	 Company}\\
	 }}

\def\abstracts#1#2#3{{
	\centering{\begin{minipage}{4.5in}\baselineskip=10pt\footnotesize
	\parindent=0pt #1\par 
	\parindent=15pt #2\par
	\parindent=15pt #3
	\end{minipage}}\par}} 

\newcommand{\bibit}{\nineit}

\renewenvironment{thebibliography}[1]
	{\frenchspacing
	 \ninerm\baselineskip=11pt
	 \begin{list}{\arabic{enumi}.}
	{\usecounter{enumi}\setlength{\parsep}{0pt}
	 \setlength{\leftmargin 12.7pt}{\rightmargin 0pt} 
	 \setlength{\itemsep}{0pt} \settowidth
	{\labelwidth}{#1.}\sloppy}}{\end{list}}

\newcounter{itemlistc}
\newcounter{romanlistc}
\newcounter{alphlistc}
\newcounter{arabiclistc}

\newcommand{\fcaption}[1]{
        \refstepcounter{figure}
        \setbox\@tempboxa = \hbox{\footnotesize Fig.~\thefigure. #1}
        \ifdim \wd\@tempboxa > 5in
           {\begin{center}
        \parbox{5in}{\footnotesize\smalllineskip Fig.~\thefigure. #1}
            \end{center}}
        \else
             {\begin{center}
             {\footnotesize Fig.~\thefigure. #1}
              \end{center}}
        \fi}

\newcommand{\tcaption}[1]{
        \refstepcounter{table}
        \setbox\@tempboxa = \hbox{\footnotesize Table~\thetable. #1}
        \ifdim \wd\@tempboxa > 5in
           {\begin{center}
        \parbox{5in}{\footnotesize\smalllineskip Table~\thetable. #1}
            \end{center}}
        \else
             {\begin{center}
             {\footnotesize Table~\thetable. #1}
              \end{center}}
        \fi}

\def\@citex[#1]#2{\if@filesw\immediate\write\@auxout
	{\string\citation{#2}}\fi
\def\@citea{}\@cite{\@for\@citeb:=#2\do
	{\@citea\def\@citea{,}\@ifundefined
	{b@\@citeb}{{\bf ?}\@warning
	{Citation `\@citeb' on page \thepage \space undefined}}
	{\csname b@\@citeb\endcsname}}}{#1}}

\newif\if@cghi
\def\cite{\@cghitrue\@ifnextchar [{\@tempswatrue
	\@citex}{\@tempswafalse\@citex[]}}
\def\citelow{\@cghifalse\@ifnextchar [{\@tempswatrue
	\@citex}{\@tempswafalse\@citex[]}}
\def\@cite#1#2{{$\null^{#1}$\if@tempswa\typeout
	{IJCGA warning: optional citation argument 
	ignored: `#2'} \fi}}

\def\pmb#1{\setbox0=\hbox{#1}
	\kern-.025em\copy0\kern-\wd0
	\kern.05em\copy0\kern-\wd0
	\kern-.025em\raise.0433em\box0}

\def\fnt#1#2{\footnotetext{\kern-.3em
	{$^{\mbox{\scriptsize #1}}$}{#2}}}

\def\fpage#1{\begingroup
\voffset=.3in
\thispagestyle{empty}\begin{table}[b]\centerline{\footnotesize #1}
	\end{table}\endgroup}

\def\runninghead#1#2{\pagestyle{myheadings}
\markboth{{\protect\footnotesize\it{\quad #1}}\hfill}
{\hfill{\protect\footnotesize\it{#2\quad}}}}
\font\tenrm=cmr10
\font\tenit=cmti10 
\font\tenbf=cmbx10
\font\bfit=cmbxti10 at 10pt
\font\ninerm=cmr9
\font\nineit=cmti9

\font\eightrm=cmr8

\def\qed{\hbox{${\vcenter{\vbox{			
   \hrule height 0.4pt\hbox{\vrule width 0.4pt height 6pt
   \kern5pt\vrule width 0.4pt}\hrule height 0.4pt}}}$}}

\renewcommand{\thefootnote}{\fnsymbol{footnote}}	

\begin{document}

\runninghead{Model Independent Extractions of $|V_{ub}|$ From
Inclusive $B$ Decays} {Model Independent 
Extractions of $|V_{ub}|$ From Inclusive $B$ Decays}

\normalsize\textlineskip
\thispagestyle{empty}
\setcounter{page}{1}

\copyrightheading{}			

\vspace*{0.88truein}

\fpage{1}
\centerline{\bf MODEL INDEPENDENT EXTRACTIONS OF $|V_{ub}|$ FROM}
\vspace*{0.035truein}
\centerline{\bf INCLUSIVE $B$ DECAYS}
\vspace*{0.37truein}
\centerline{\footnotesize IAN LOW
}
\vspace*{0.015truein}
\centerline{\footnotesize\it Department of Physics, Carnegie Mellon 
University, Pittsburgh, PA 15213}
\baselineskip=10pt
\vspace*{0.225truein}
\vspace*{0.21truein}
\abstracts{
We discuss the possibility of extracting $|V_{ub}|$ from various spectra
of inclusive $B$ decays, without large systematic errors which usually
arise from having to model the Fermi motion of the heavy quark. This
strategy can be applied to the electron energy spectrum, as well as
the hadronic mass spectrum.
Modulo violation of local hadron-parton duality, the theoretical error in
the extraction is estimated to be less than 10\%.}{}{}
\textlineskip			
\vspace*{12pt}			


Among the nine entries of the Cabbibo-Kobayashi-Maskawa (CKM) matrix, 
$V_{ub}$ and $V_{td}$ are the most important
elements for understanding CP violation. Unfortunately, these two 
elements are
the most difficult to extract experimentally. 
When making theoretical
predictions, one is often forced to 
resort to various models
 due to our inability to
calculate hadronic dynamics from QCD. The trouble with models is that
we have no idea what the theoretical uncertainty is.
In what follows I will describe a method of extracting
 $|V_{ub}|$ from the inclusive
semileptonic decays $B\to X_u\ell\nu$ without having to model hadronic
dynamics.

\textheight=7.8truein
\setcounter{footnote}{0}
\renewcommand{\thefootnote}{\alph{footnote}}

In the standard model the decay of $B$ mesons can be calculated in a 
systematic expansion in $\alpha_s$, 
$1/(M_W,m_t)$ and $1/m_b$. First the full standard model is matched 
onto the four fermion theory. It is then run down to the scale $m_b$, after
which the decay rate may be calculated in heavy quark
effective theory (HQET) in expansion in $1/m_b$. Using the optical
theorem, the inclusive decay rate can be computed by taking the imaginary
part of the forward scattering amplitude and integrating along 
the physical cut in the complex $v\cdot q$ plane, where $q$ is the momentum
transfered to the lepton pair and $v$ is the four velocity of the $B$ meson.
The integration contour can then be deformed away from the resonance region
into a region where perturbative calculation can be trusted.$^1$ Therefore,
a hadronic quantity averaged over a sufficient number of states can
be computed reliably using QCD perturbation theory.
This is the justification of (global) parton-hadron duality.

Unfortunately, at the present stage we are unable to measure the total
decay rate of $b\to u\ell\nu$ due to the overwhelming charmed
background. It is thus necessary to introduce experimental cuts
on various differential spectra. The traditional method is to put a cut
on the lepton energy, $M_B/2
> E_\ell > (M_B^2-M_D^2)/(2M_B)$, which leaves us with a window
of $2\Delta E/m_b \approx 0.13$. It is also possible to use the cut
hadronic invariant mass spectrum$^2$ by considering hadronic final states
with invariant mass $M_D^2> s_H>0$, leaving a window of $\Delta s_H/m_b^2 
\approx 0.15$. Recently it has been proposed that one can also use 
the cut leptonic invariant mass spectrum$^3$ $M_D^2 > q^2 > (M_B-M_D)^2$, 
with a window of $\Delta q^2/m_b^2 \approx 0.57$. In any case, the
necessity of experimental cut introduces a third
scale $\rho$, in addition to $m_b$ and $\Lambda_{\rm{QCD}}$,
which measures the relative size of the phase space of interest.

To see the effect of $\rho$ on the theoretical calculation, let us
consider the end-point region of the lepton energy spectrum. At
${\cal O}(\alpha_s)$ the spectrum contains infrared and collinear
logarithms like $\log^2(1-x)$
and $\log(1-x)$, where $x=2E_\ell/m_b$. In the end-point region,
$\alpha_s(m_b) \log^2(1-\rho) \approx 0.8$, which signifies the poor 
convergence of QCD perturbation theory. 
One therefore needs to resum logarithms
of the form $\alpha_s^m \log^n(1-x)$. 
As for the non-perturbative expansion in $1/m_b$, 
it has singular terms$^4$ $\delta(1-x)$ and
$\delta^\prime(1-x)$. This suggests that the relevant
expansion parameter in this region is ${\Lambda/[m_b(1-\rho)]}
\sim {\cal O}(1)$ and one
needs to take into account an infinite number of terms 
like $\delta^{(n)}(1-x)$
all contributing at leading order.

To achieve the above goals, it is useful to utilize the infrared 
factorization in the lepton energy end-point spectrum.$^5$ It can be shown
that, when $x \to 1$, $s_H \sim {\cal O}(m_b(1-x))$ and 
$E_H \sim {\cal O}(m_b)$. The $u$-quark produces a {\it jet} of 
collinear particles whose invariant mass approaches zero with total
energy held fixed. In addition, the jet hadronizes at a much later time
in the rest frame of the $B$ meson, due to the time dilation. The 
differential decay rate thus factorizes into three parts, the hard,
the jet, and the soft, characterized by three disparate scales:
${\cal O}(m_b)$, ${\cal O}(m_b\sqrt{1-x})$, and ${\cal O}(m_b(1-x))$
respectively,
\begin{equation}
\left. \frac{d^3\Gamma}{dxdq^2d(v\cdot q)} \right|_{x\to 1} \sim
\int dz\ S(z,\mu) J(z,\mu) H(\mu),
\end{equation}
where $\mu$ is the factorization scale. Physical amplitudes should 
not depend on $\mu$ and this leads to a renormalization
group equation which can be used to resum the infrared and
collinear logs.$^{5,6}$ Taking the moment of the spectrum further
leads to
\begin{eqnarray}
M_N &=& - \frac1{\Gamma_0} \int_0^{M_B/m_b} dx\ x^{N-1} \frac{d}{dx}
         \frac{d\Gamma}{dx} \nonumber\\
    &=& f_N \sigma_N J_N H_N + {\cal O}(1/N) 
\end{eqnarray}
$H_N$ contains the short distance interaction at scale $m_b$ and 
can be computed perturbatively. $\sigma_N$ and $J_N$ contain
the infrared and collinear logs, which are to be resummed using RG equation.
$f_N$ contains interaction at scale $\Lambda_{\rm QCD}$ and is strictly
non-perturbative. A formal expression for $f_N$ in the $x$-space is
\begin{equation}
f(k_+) = \langle B(v)|\ \bar{b}_v\ \delta(k_+ -iD_+)\ b_v\ |B(v)\rangle,
\end{equation}
which is the light-cone distribution function of the $b$-quark inside
the $B$ meson and resums the most singular terms like $\delta^{(n)}(1-x)$
in HQET.$^{7}$ Since $f(k_+)$ contains the long distance physics and is 
not calculable from QCD, one typically has to resort to modeling. On the
other hand, this implies that $f(k_+)$ is insensitive to short
distance physics and thus universal in inclusive $B$ decays. 
One can extract $f(k_+)$ from the inclusive 
$B\to X_s\gamma$ decays. In the end-point region of photon spectrum
similar infrared factorization holds and the moment factorizes into$^{5,8}$
\begin{eqnarray}
M_N^\gamma &=& -\frac1{\Gamma_0^\gamma}\int_0^{M_B/m_b} dx^\gamma
  (x^\gamma)^{N-1} \frac{d\Gamma^\gamma}{dx^\gamma} \nonumber \\
   &=& f_N \sigma_N J_N^\gamma H_N^\gamma
\end{eqnarray}
Replace $f_N \sigma_N$ in (2) by $M_N^\gamma$ using (4) and take the
inverse Mellin transform to go back to $x$-space we obtain$^{9}$
\begin{equation}
\frac{|V_{ub}|^2}{|V_{ts}^*V_{tb}|^2} \sim
\frac{\delta\Gamma(B\to X_u\ell\nu, \rho)}{\int\int_\rho 
\frac{d\Gamma^\gamma}{dx^\gamma} * K(x^\gamma;\alpha_s)},
\end{equation}
where the kernel $K(x^\gamma;\alpha_s)$ can be computed 
perturbatively.
Corrections to this are ${\cal O}(\Lambda/m_b, \alpha_s(1-\rho),(1-\rho)^3)$.
These corrections are of order 10\% in $|V_{ub}|^2$.

However, this assumes that parton-hadron duality works well. In fact,
the end-point region of the electron energy spectrum contains only about
10\% of the total rate, whereas the low $s_H$ region of the hadronic
invariant mass spectrum contains 40-80\% of the total rate. One therefore
expects that parton-hadron duality works better in this case. A similar
strategy to extract $|V_{ub}|$ can be applied.$^{10}$   The leptonic
invariant mass spectrum might be a theoretically clean method$^3$, but
the cut contains only 20\% of the total rate. In the end it is clear
that no single extraction of $|V_{ub}|$ from inclusive decays should
be trusted. We would have faith only after convergence among several
independent extractions. Comparision among different extractions
would also shed some light on the validity of parton-hadron duality.

{\it Acknowledgements}: 
I would like to thank Adam Leibovich and Ira Rothstein for
collaboration on this topic and Mike Luke and Matthias Neubert for helpful
comments. 
This work was supported in part by the Department of
Energy under grant number DOE-ER-4068-143.



\noindent
\begin{thebibliography}{000}

\bibitem{1}
J. Chay, H. Georgi, and B. Grinstein, {\bibit Phys. Lett. B247}
(1990) 399.

\bibitem{2}
V. Barger, C.S. Kim, and R.J.N. Phillips, {\bibit Phys. Lett. B251} (1990)
629; J. Dai, {\bibit Phys. Lett. B333} (1994) 212; C. Greub and S.J. Rey,
{\bibit Phys. Rev. D56} (1997) 4250; A. Falk, Z. Ligeti and M.B. Wise, 
{\bibit Phys. Lett. B406} (1997) 225.

\bibitem{3}
C.W. Bauer, Z. Ligeti and M. Luke, {\bibit Phys. Lett. B479} (2000) 395.

\bibitem{4}
A.V. Manohar and M.B. Wise, {\bibit Phys. Rev. D49} (1994) 1310;
B. Blok, L. Koyrakh, M. Shifman and A.I. Vainshtein, {\bibit Phys. Rev.
D49} (1994) 3356.

\bibitem{5}
G. Sterman and G.P. Korchemsky, {\bibit Phys. lett. B340} (1994) 96.

\bibitem{6}
R. Akhoury and I.Z. Rothstein, {\bibit Phys. Rev. D54} (1996) 2349.

\bibitem{7}
M. Neubert, {\bibit Phys. Rev. D49} (1994) 3392; T. Mannel and M. Neubert
{\bibit Phys. Rev. D50} (1994) 2037; I.I. Bigi, M.A. Shifman, 
N.G. Uraltsev, and A.I. Vainshtein, {\bibit Int. J. Mod. Phys. A9} 
(1994) 2467. 

\bibitem{8}
A.K. Leibovich and I.Z. Rothstein, {\bibit Phys. Rev. D61} (2000)
074006.

\bibitem{9}
A.K. Leibovich, I. Low and I.Z. Rothstein, {\bibit Phys. Rev. D61}
(2000) 053006.

\bibitem{10}
A.K. Leibovich, I. Low and I.Z. Rothstein, {\bibit Phys. Rev. D62}
(2000) 014010; \\
A.K. Leibovich, I. Low and I.Z. Rothstein, {\bibit Phys. Lett. B486}
(2000) 86.






\end{thebibliography}
\end{document}